# Plasmon hybridization in pyramidal metamaterials: a route towards ultra-broadband absorption

**Michaël Lobet,[1] Mercy Lard,[2] Michaël Sarrazin,[1] Olivier Deparis,[1] Luc Henrard[1,*]**

[1]*Solid-State Physics laboratory, Department of Physics, University of Namur, 61 rue de Bruxelles, 5000 Namur, Belgium*
[2]*Nanometer Structure Consortium (nmC@LU) and Solid State Physics Lund University, Box 118, SE-221 00 Lund, Sweden*
*\* luc.henrard@unamur.be*

**Abstract**: Pyramidal metamaterials are currently developed for ultra-broadband absorbers. They consist of periodic arrays of alternating metal/dielectric layers forming truncated square-based pyramids. The metallic layers of increasing lengths play the role of vertically and, to a less extent, laterally coupled plasmonic resonators. Based on detailed numerical simulations, we demonstrate that plasmon hybridization between such resonators helps in achieving ultra-broadband absorption. The dipolar modes of individual resonators are shown to be prominent in the electromagnetic coupling mechanism. Lateral coupling between adjacent pyramids and vertical coupling between alternating layers are proven to be key parameters for tuning of plasmon hybridization. Following optimization, the operational bandwidth of Au/Ge pyramids, i.e. the bandwidth within which absorption is higher than 90%, extends over a 0.2-5.8 µm wavelength range, i.e. from UV-visible to mid-infrared, and total absorption (integrated over the operational bandwidth) amounts to 98.0%. The omni-directional and polarization-independent high-absorption properties of the device are verified. Moreover, we show that the choice of the dielectric layer material (Si versus Ge) is not critical for achieving ultra-broadband characteristics, which confers versatility for both design and fabrication. Realistic fabrication scenarios are briefly discussed. This plasmon hybridization route could be useful in developing photothermal devices, thermal emitters or shielding devices that dissimulate objects from near infrared detectors.

## 1. Introduction

Metamaterials have provided unprecedented ways to manipulate light at the nanoscale by engineering electromagnetic properties of materials that are not available in Nature. Metamaterials having a negative index of refraction [1] or invisibility cloaks [2,3] are some examples of those new functionalities achievable by tailoring the electric and magnetic responses. In most cases, however, metal losses set crucial limitations on devices. On the other hand, regarding nearly total (100%) absorption at a specific wavelength as an opportunity [4,5], the achievement of total absorption over a broad spectral range becomes a challenge. For instance, collecting as much solar energy as possible is of prime necessity for photothermal and photovoltaic devices. Total absorption of incoming radiation is also useful to hide objects from radar detection or infrared detectors [6]. Moreover, in integrated opto-electronics, preventing crosstalk between optical interconnects is a crucial issue that can be solved by such metamaterial devices [5,6]. Metamaterials made of metal-dielectric structures recently showed interesting properties in the context of broadband absorption [7-17]. For example, a stack of metal (Au) and nanocomposite (Au/SiO$_2$) layers has been used in order to efficiently absorb light in the visible spectrum [7,8]. Double-layered metallic nanostructures surrounded by dielectric medium were shown to have narrow resonance peaks in the absorption spectrum [9,10]. Recently, saw-toothed anisotropic metamaterials were proposed for broadband absorption in the infrared [14], microwave [15] or visible [17] ranges with angular selective thermal emission properties [16].

The present work exploits the concept of multilayer Au/Ge pyramidal structures [14,17] in order to demonstrate the theoretical possibility to greatly extend the spectral range over which nearly total absorption (> 90%) is achieved. We obtain here an ultra-broadband operation ranging from 200 nm to 5.8 μm is obtained, encompassing the UV-visible, the near-infrared and mid-infrared regions. Plasmon hybridization of dipolar modes of coupled layered resonators is revealed to be responsible for the ultra-broadband characteristic. Versatility in the choice of the dielectric spacer material is highlighted as well as omni-directionality of the absorption.

This article is organized as follows. First, general considerations related to the design of an efficient metamaterial absorber (MMA) are presented on the basis of a three-dimensional generalization of a previously reported structure [14] (section 2). Then, in section 3, the layer-by-layer building of the MMA is analyzed in order to understand the absorption mechanism. Numerical simulations are used to confirm that electromagnetic coupling is responsible for the ultra-broadband property and to identify optimal structure parameters. In section 4, through the study of localized surface plasmons (LSP) on dielectric-metal-dielectric slabs and their spectral aggregation via electromagnetic coupling, we show that the origin of the absorption spectral broadening is related to size-dependent LSP excitations. The relative insensitivity of the device performance to the chosen dielectric material is demonstrated. Finally, possible practical implementation is discussed.

## 2. Principle and design of the metamaterial absorber

Light impinging on a metamaterial can be transmitted, reflected or absorbed. Transmittance $T(\lambda)$, reflectance $R(\lambda)$ and absorption $A(\lambda)$, are related by the energy conservation law:

$$T(\lambda) + R(\lambda) + A(\lambda) = 1. \tag{1}$$

In order to mimic a perfect black body, one must block transmission $(T(\lambda)=0)$. High absorption $A(\lambda)$ requires low reflectance $R(\lambda)$ and efficient energy dissipation through excitation of the eigenmodes of the structure. Anti-reflection are well known devices [18-20], in which surface corrugation provides a gradual transition of refractive index between air and dielectric material, hence effective suppression of reflection over a broadband spectral range. Combination of efficient anti-reflection property with coupled plasmonic resonators is commonly used to reach nearly perfect absorption over a broadband range of wavelengths [14-17].

The structure of the investigated MMA is shown in Fig.1(a). For the purpose of blocking transmission, the MMA is placed on a 200-nm thick gold film on top of a dielectric slab (SiO$_2$, refractive index $n_{SiO2}$ = 1.44) acting as semi-infinite substrate. This gold film, in addition to blocking transmission, acts as a mirror in the infrared, enabling light to re-interact with the dedicated absorbing structure on top. The MMA structure consists of a square array of truncated pyramids with a squared base as proposed in [14-17]. Our starting structure is a generalization in three dimensions of the structure reported in [14]. It has *N = 20* layers with alternating gold layers (thickness $t_{Au}$ = 15 nm) and dielectric layers (thickness $t_{diel}$ = 35 nm) with a linear lateral side reduction from *L= 600* nm at the basis to *l= 150* nm at the top of the pyramid. The total thickness is therefore *t = 1* μm. The gold thickness is

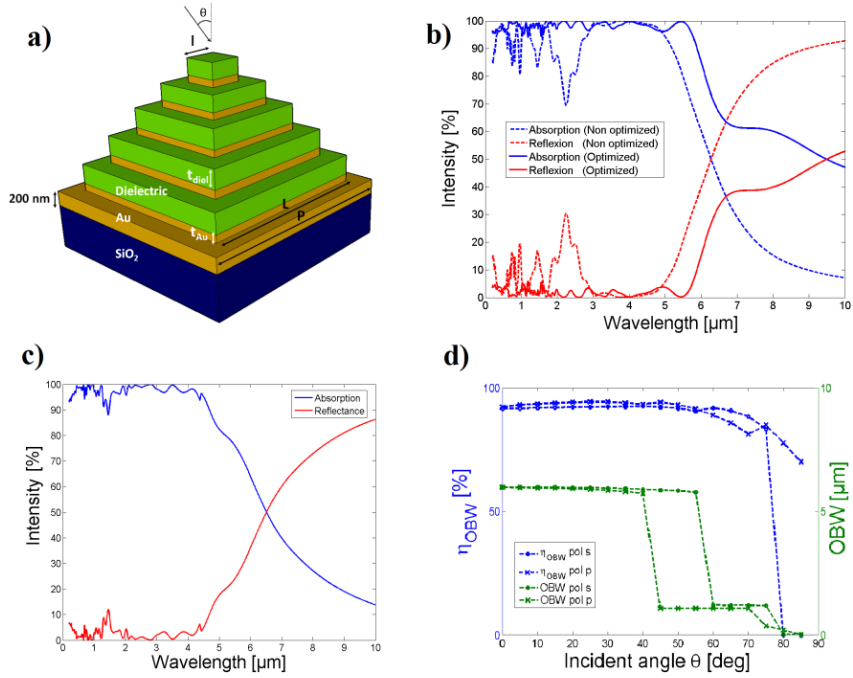

Fig. 1 (a) Metamaterial absorber (MMA) made of truncated pyramids forming a square array of period *P*, (b) Absorption (blue) and reflectance (red) spectra of the MMA (non optimized structure: dashed line, optimized structure: solid line) with Ge as dielectric spacer under normally incident illumination, (c) Absorption (blue) and reflectance (red) spectra of the MMA with Si as dielectric spacer under normally incident illumination, (d) Angular dependencies of $\eta_{OBW}$ (left axis, blue lines) and OBW (right axis, green lines) of the MMA for s-polarized (square) and p-polarized (dot) incident light.

chosen to be less than the skin depth over the considered wavelength range in order to enable coupling of surface plasmons between both sides of the gold square layers as later explained. The lateral period of the starting structure is set to *P = 800* nm. The influence of the number of layers, dielectric thickness and lateral period will be studied and these parameters will be optimized.

Each metallic layer of the pyramid can be viewed as a plasmonic resonator. In order to make an efficient plasmonic resonator, a metal with high conductivity is required [21]. Moreover, the pyramidal shape is of prime necessity in order to minimize the reflectance of the MMA: the graded refractive index due to the pyramidal metal-dielectric layer arrangement acts as an anti-reflective device. We chose gold as the metal and germanium as the dielectric spacer in order to fulfill the above conditions, although other combinations of materials could also be used. Complex refractive indices of Au and Ge are described using tabulated data [22, 23].

## 3. Device performances

The figure of merit $\eta_{OBW}$ is first defined in order to characterize the performances of the MMA. It is based on the operational bandwidth $OBW = \lambda_f - \lambda_i$, i.e. the spectral region where absorption is higher than a threshold value, and is defined as the absorption integrated over the operational bandwidth:

$$\eta_{OBW} = \frac{\int_{\lambda_i}^{\lambda_f} A(\lambda) d\lambda}{\lambda_f - \lambda_i}, \quad (2)$$

where $\lambda_i = 0.2$ μm is fixed and $\lambda_f$ is determined by the threshold absorption condition. Here, we first set the threshold at $A(\lambda) > 70\%$ for all wavelengths.

Numerical simulations were performed in order to calculate the absorption spectrum of the MMA. The lateral periodicity and the stratified geometry of the MMA structure are perfectly suitable for three-dimensional scattering-matrix calculation methods. The Rigorous Coupled Wave Analysis (RCWA) method is chosen here for solving Maxwell equations [24, 25]. Reflectance and transmittance are calculated from the Poynting vector in incidence and emergence media, respectively. The strong refractive index mismatch between gold, dielectric medium (Ge) and surrounding air is critical from a numerical accuracy point of view. For this reason, we used 19 x 19 plane waves (along $x$ and $y$ directions respectively) in the Fourier expansions as a requirement to ensure numerical convergence within reasonable calculation time. We note that lateral coupling between adjacent pyramids and vertical coupling between metallic layers of a single pyramid are fully taken into account in RCWA simulations.

Before any optimization, reflectance and absorption spectra of the MMA were calculated for normally incident ($\theta = 0°$, $\varphi = 0°$) light and the geometrical parameter values described in the previous section (Fig. 1(b), dashed lines). The absorption spectrum of this starting (non-optimized) structure exhibits a high and flat absorption, especially between 3 μm and 5 μm. The operational bandwidth extends between $\lambda_i = 0.2$ μm and $\lambda_f = 5.7$ μm, and the figure of merit is $\eta_{OBW} = 92.2\%$. This operational bandwidth of $OBW=5.5$μm is a factor of two improvement over recently reported values [14,17] due to the full consideration of the three dimensional structure. At longer wavelengths, the absorption linearly decreases from 70 % (at 5.7 μm), to 7.2% (at 10 μ$m$). Moreover, the broadband absorption property of the proposed MMA is almost independent of the angle of incidence and polarization state (Fig. 1(d)). The merit factor $\eta_{OBW}$ maintains significantly higher values at any incident angle $\theta$ lower than 75°. An angle-averaged value can be calculated for incident angles between $\theta = 0°$ and $\theta = 75°$ and is found to be equal to $\bar{\eta}_{OBW} = 91.1\%$. The OBW remains accordingly constant in the angle range 0°-55°. The sharp drop of OBW at 60° corresponds to the more intense reflectance peak at 2.2 μm (Fig. 1(b), dashed red line) with increasing angle, absorption at this wavelength falls below the threshold value (70%) as the incident angle reaches ≈60°. At more grazing incident angles, reflectance increases as predicted by Fresnel coefficients expressions, resulting in a lower absorption, hence a reduced OBW. Simulations were performed with both s-polarized and p-polarized incident light at various incident angles. The MMA performances are found to be fairly independent of the incident light polarization (cf. discussion, §4).

For a better understanding of the absorption mechanism, we studied, layer-by-layer, the electromagnetic response of the MMA, by building the structure from the bottom to the top of the pyramid. This methodology aided in understanding the influence of each additional layer acting on the absorption spectrum. It also provided information about the number of layers required to reach better absorption properties with respect to a threshold. For this optimization, we count the number of plasmonic resonators $N$, from bottom *(L= 600 nm, N = 1)* to top *(l = 75 nm, N = 23)*. The lateral period is unchanged. The absorption increases progressively as the pyramid is built due to a growing number of resonance peaks appearing in the spectrum as the bandwidth is broadened (Fig. 2). Above *N = 20 (l = 150 nm)*, the absorption does not increase further, leading to a saturation regime: we will then consider *N = 20* in the following of the study. Additionally, a global blue shift of the absorption peaks is noticed when increasing the number of layers, the origin of which will be discussed later.

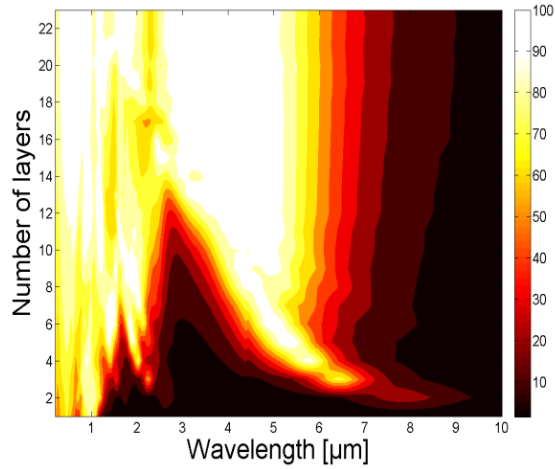

Fig 2. Absorption map of the MMA as a function of the number of layers added from bottom ($L = 600$ nm) ($L = 600\ nm$) to top ($l = 75$ nm). Lattice period is fixed to $P = 800$ nm. Adding layers enlarges the operational bandwidth through additional plasmonic resonators which couple to each other, resulting in a blue shift of individual plasmonic modes.

We further optimized two other geometrical parameters which may influence the coupling between resonators and hence the OBW: namely, the lateral period $P$ (lateral coupling) and the thickness of dielectric spacers $t_{diel}$ (vertical coupling). Regarding the lateral period (Fig. 3), both the OBW and $\eta_{OBW}$ reach maximum values at $P = 650$ nm. For larger periods, the absorption linearly drops and the ultra-broadband characteristic is progressively lost .These results confirm that coupled stacked resonators do interact laterally.

Using $P = 650$ nm which maximizes both OBW and $\eta_{OBW}$ we then optimized the dielectric thickness. As shown in Fig. 4, the absorption is quite insensitive to the dielectric spacer thickness in a specific range (thickness between 25 nm and 200 nm). Above a thickness of 200 nm, the absorption drops in the 1-2 μm wavelength range. The figure of merit $\eta_{OBW}$ peaks at 96.6% for a dielectric thickness equal to 175 nm. The OBW has a maximum value of 6.9 μm for a dielectric thickness of 225 nm but drops sharply above this thickness value due to lowered absorption in the 1-2 μm wavelength range. Our optimal value choice ($t_{diel}$ = 175 nm) is not as stringent as for the lateral period: any value between 25 nm and 225 nm could be used, depending on the device application. However, we note that absorption, in the wavelength range above 6 μm, exhibits higher values than those of the non-optimized MMA (Figs. 2 and 3). Increasing the dielectric thickness positively influences absorption at longer wavelengths. In summary, the optimization leads to the following optimal values: number of layer $N = 20$, lateral period $P = 650$ nm and dielectric spacer thickness $t_{diel}$ = 175 nm. The resulting absorption spectrum is shown in Fig. 1(b) for the optimized MMA (solid line). Thanks to the optimization, we are able to increase the absorption threshold value, and we therefore fix it to 90%. The operational bandwidth, over which absorption is higher than 90%, is then equal to 5.6 μm over the 0.2-5.8 μm spectral range. The figure of merit is now $\eta_{OBW,90} = 98.0\%$. This is even better than other structures having OBW in the visible and near infrared ranges [7,8,11,17].

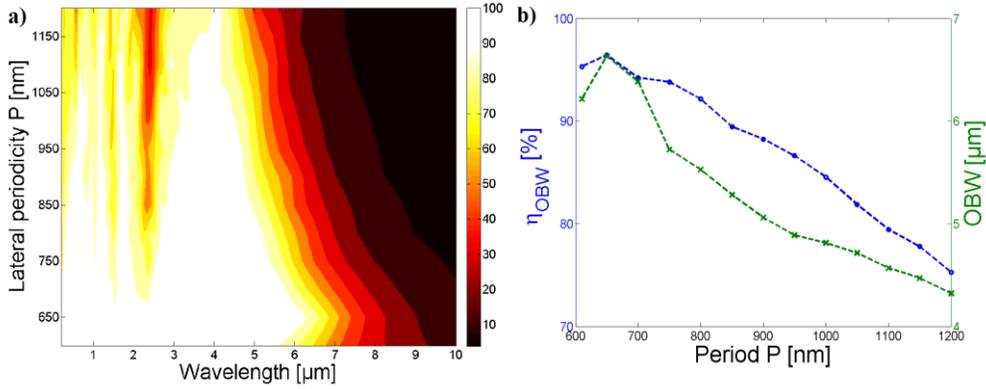

Fig. 3 (a) Absorption spectrum of the MMA as function of lateral period *P* increasing from 610 nm to 1200 nm. (b) Operational bandwidth *OBW* (right axis, green crosses) and figure of merit $\eta_{OBW}$ with a threshold value of 70% (left axis, blue dots) as a function of lateral period. $\eta_{OBW}$ shows a maximum value of 96.4 % for *P = 650* nm. The OBW shows a maximum of 6.6 µm for the same lateral period.

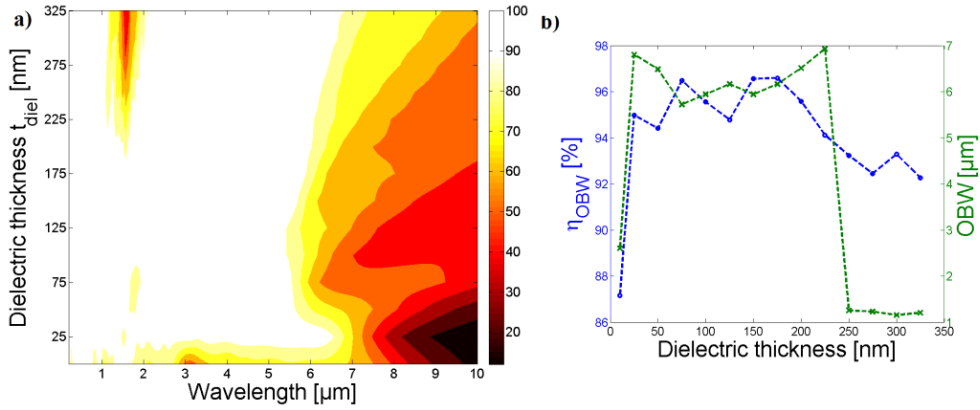

Fig. 4 (a) Absorption spectrum of the MMA as a function of the dielectric spacer thickness increasing from 10 nm to 325 nm. The MMA is composed of *N= 20* alternating gold and germanium layers and has a lateral period *P* = 650 nm. (b) Operational bandwidth *OBW* (right axis, green crosses) and figure of merit $\eta_{OBW}$ with a threshold value of 70% (left axis, blue dots) as a function of dielectric thickness. $\eta_{OBW}$ shows a maximum value of 96.6 % for a dielectric thickness of 175 nm. The OBW shows a maximum of 6.9 µm for a dielectric thickness of 225 nm.

### 4. Discussion

Hereafter, we aim at explaining the broadband absorption mechanism, the observed blue shift of the absorption spectrum with increasing number of layers and the influence of the lateral and vertical coupling. The ultra-broadband property of the proposed MMA arises from the collection of coupled plasmonic resonances whose narrow absorption peaks sum up to form a broad spectrum, as suggested in section 3 and in previous studies [14,17]. The resonators consist of the individual square gold layers of the pyramid, in which the incident wave excites localized surface plasmon modes at the gold-dielectric interfaces. Since surface plasmons are shape and size dependent, by changing the dimensions of the resonators, the resonances of each layer are located at different frequencies, resulting in the broadening of the absorption spectrum of the whole structure [14,17]. This behavior was illustrated in Fig. 2 by building the MMA layer-by-layer. In addition, the increasing number of coupled resonators in the pyramidal structure enables the tuning of the vertical coupling and causes a blue shift of the individual plasmonic resonances (Fig. 4).

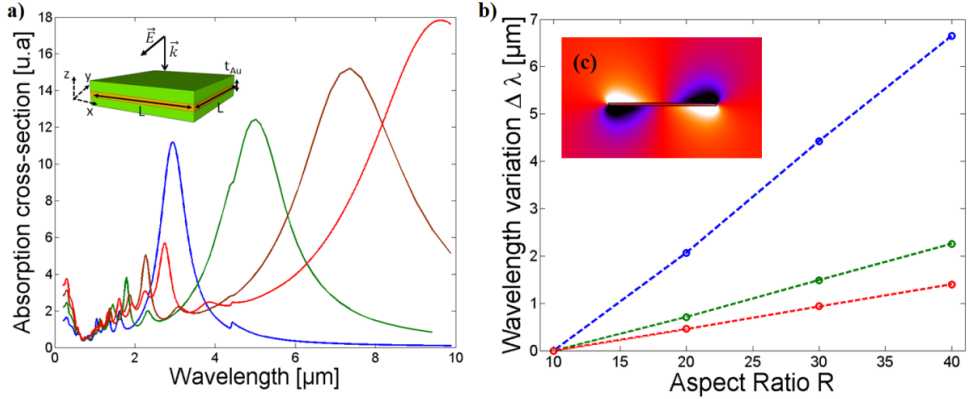

Fig 5. (a) Absorption cross-section spectra of square gold layers of lengths *L*: DDA simulations using $N_d = 1 \times 10^5$ dipoles (L = 150 nm, blue line), $N_d = 4 \times 10^5$ dipoles (L = 300 nm, green line), $N_d = 1.125 \times 10^5$ dipoles (L = 450 nm, brown line) and $N_d = 2 \times 10^5$ dipoles (L = 600 nm, red line). The interdistance between dipoles is dx = dy = dz = 3 nm (b) Variations of LSP wavelengths with aspect ratio $R = L/t_{Au}$ for dipolar mode (blue), sextupolar mode (green) and octupolar mode (red). (c) Field distribution (imaginary part of $E_z$) in y-z plane of the dipolar mode of the $L = 600\ nm$ square gold layer in germanium surrounding medium.

Let us first consider the electromagnetic response of individual layers. The absorption cross-sections of individual 15-nm thick square gold layers embedded in infinite germanium surrounding medium were calculated for different lengths *L* of the square gold layer (Fig. 5(a)) using a discrete dipole approximation (DDA) method as implemented in DDSCAT code [30]. The DDA method enables the calculation of scattering and absorption cross sections of arbitrarily shaped particles which can be isolated, in contrast with the RCWA method where a periodic arrangement of particles is required. In DDA simulations, the number of dipoles $N_d$ and the interdistance between dipoles are the key factors for convergence (see caption of Fig. 5 for details). We note that germanium (infinite surrounding medium) is taken here as non-dispersive with a dielectric constant $\varepsilon_{Ge} = 16$. Light is normally incident with the electric field polarized in the *y*-direction (see inset, Fig. 5(a)). For all absorption cross-section spectra corresponding to different square lengths, a multi-peak structure is found and a red shift of the peaks with increasing *L* is observed. Those peaks correspond to excitation of different localized surface plasmon modes, as shown on Fig. 6 for the three most important peaks. On the *x-y* plane distributions of the imaginary part of the *z* component of the electric field, we can identify a dipolar mode (at the longest wavelength, Fig. 6(a)), and both sextupolar and octupolar modes (at shorter wavelengths, Fig. 6(b) and 6(c)) in analogy to spherical particles. The other modes, which appear in the absorption cross-section spectrum at even shorter wavelengths, are characterized by standing wave field patterns in the *y*-direction but will not be analyzed here since they have relatively lower cross-sections. The dipolar modes are identified to be the most prominent in absorption cross-sections (Fig. 5(a)). They are therefore responsible for the dominating features in the absorption spectrum of the MMA between 2 and 6μm.

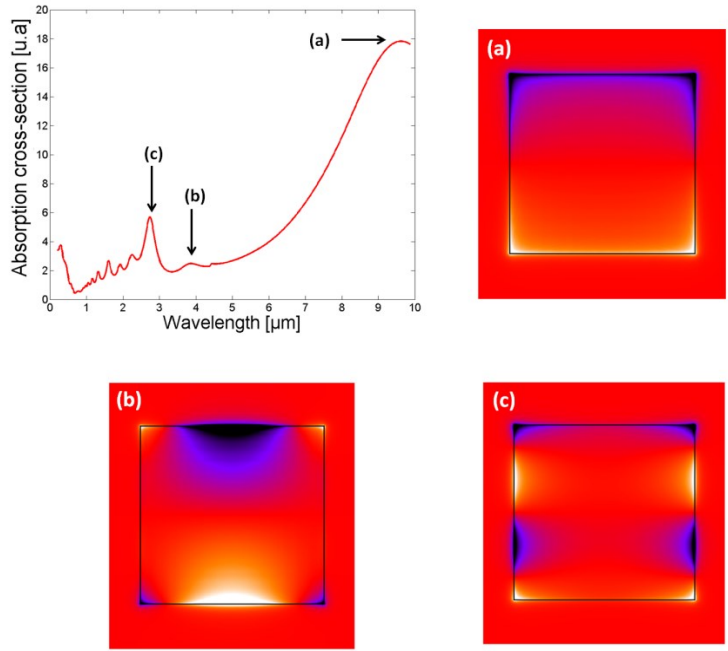

Fig. 6 Absorption cross section spectrum (top left) and field distribution (imaginary part of $E_z$) in the *x-y* plane of the three major peaks (a: dipolar mode, b: sextupolar mode, c: octupolar mode) of an individual square gold layer ($L = 600$ nm) in germanium surrounding medium.

It is known that LSP modes are dependent on the aspect ratio of the scattering object [31]. Fig. 5(b) shows the red shift of the three main modes when the aspect ratio $R = L / t_{Au}$ increases. Wavelength variation $\Delta\lambda$ is defined as the difference between the resonance wavelengths of the mode at a given aspect ratio and at the smallest investigated aspect ratio ($R = 10$). The observed linear dependence of the LSP wavelength on aspect ratio $R$ is similar to that observed for nanotriangle platelets [31] and allows for a prediction of the main absorption peak for arbitrary gold square aspect ratio. The LSPs have a symmetric charge distribution resulting from the strong coupling between the upper and lower surfaces. Following reference [31], a direct link can be made between the LSP and the surface plasmon polariton (SPP) of an infinitely thin Au film embedded in a dielectric medium. Indeed, the LSP can be viewed as the SPP with a quantified wavelength ($L = n\lambda/2$). Thus the linear $\Delta\lambda$-$R$ curve (Fig. 5(b)) is consistent with the linear dispersion relation of the SPP according to $qt_{Au}$, $q$ being the wave vector ($q = 2\pi/\lambda$). Moreover, based on the above analogy, we can predict that, if the thickness $t_{Au}$ is modified, the length $L$ has to be scaled accordingly to keep the same characteristics of the absorption. Consequently, a modification of the wavelength range of the MMA can be obtained by a change in the aspect ratio of the metallic layer, as exemplified in reference [17] for absorption over the solar spectrum. Finally, the significant enhancement of the local electromagnetic field (Fig. 6) is characteristic of LSP resonances. Particularly, we note that the electric field of the dipolar mode is enhanced at the gold-germanium interface, with a non-negligible extension on the Ge side (Fig. 5(c)).

As a matter of fact, the MMA absorption spectrum cannot be viewed exclusively as the sum of all the contributions of individual gold layers of different sizes. For example, the plasmonic dipolar mode of the $L = 600$ nm gold square appears around $\lambda = 10\ \mu m$ (Fig. 6), but no absorption occurs at that wavelength in the pyramidal structure (Fig. 1(b)). In a layered structure, plasmon modes of individual layers hybridize to form hybrid plasmon modes similarly to molecular orbital formation from single atom wave functions [26-29]. Here, both lateral and vertical coupling occurs (as revealed by Figs. 3 and 4). For the pyramidal structure, coupling between adjacent layers gives rise to the quasi-continuous absorption spectrum. This aggregation induces several dips in the reflectance spectrum over the OBW (Fig. 1(b)). Comparable results are found with similar structures made of two gold rectangular layers in the literature [28]. For a pair of identical metallic squares, the layer-to-layer coupling lifts the degeneracy of the individual plasmon modes and creates symmetric and antisymmetric plasmon modes. The total dipolar moment is reinforced for the symmetric mode, creating a bright mode, while the total dipolar moment cancels out for the antisymmetric mode, creating a dark mode. Hereafter, we examine more carefully the pyramidal systems made of the first layers of the whole pyramidal structure. Figure 7 reproduces the detailed absorption spectra of Fig. 2, with the number of layers $N$ ranging from one to four. With $N = 1$ layer, the system behaves like a shallow gold grating [32] on top of thin gold mirror. It is not identical to the isolated system studied above (Fig. 5) for several reasons: the isolated system is composed of a gold square layer surrounded by infinite Ge medium while the present system (Fig. 7(a)) is composed of a periodic array of gold square layers with dielectric top layers over a thin gold film (inset Fig. 7(a)). For this system, absorption occurs at visible wavelengths and vanishes in the infrared: i.e. the system, essentially the gold film, acts as a mirror. When $N = 2$, the system closely resembles a single square layer over a thin film with shallow corrugation (inset Fig. 7(b)), and additional peaks appear in the absorption spectrum due to hybridization of plasmons. The mode appearing at 7.8 μm originates from the dipolar modes of both square layers, oscillating in phase, i.e. it is a symmetric hybridization mode. The wavelength of this resonance is blue-shifted compared to that of the isolated square layers of similar sizes (Fig. 5(a)) due to hybridization. This blue-shift intensifies as the number of layers increases further and produces the spectral aggregation of the individual resonance modes (Fig. 5), resulting in the broadband property of the device (Figs 1(b) and 2). Moreover, at the resonance frequencies of the coupled system, the electromagnetic field is strongly enhanced between adjacent square layers, allowing absorption to occur also in the dielectric spacer near the metal-dielectric interface (Fig. 5(c)), which might be interesting for photovoltaic applications.

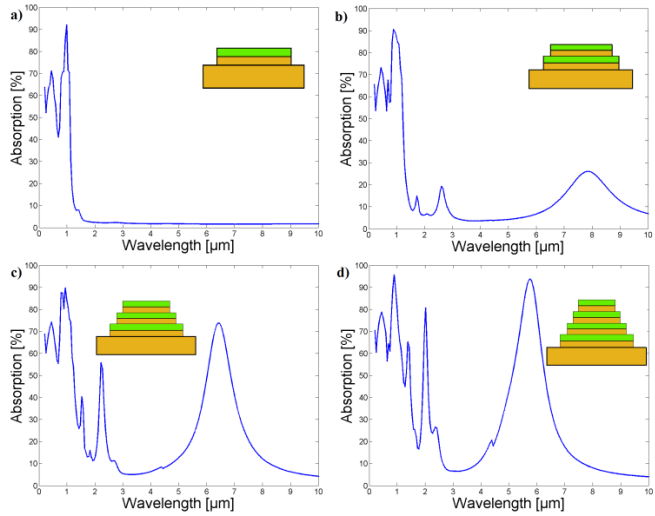

Fig. 7 Absorption spectra of pyramidal systems with a) $N= 1$, b) $N = 2$, c) $N = 3$ and d) $N = 4$ layers. For $N= 1$, the absorption spectrum is typical of a shallow grating with high reflectance in the infrared. For $N>1$, supplementary modes appear and shift to the blue due to plasmon hybridization.

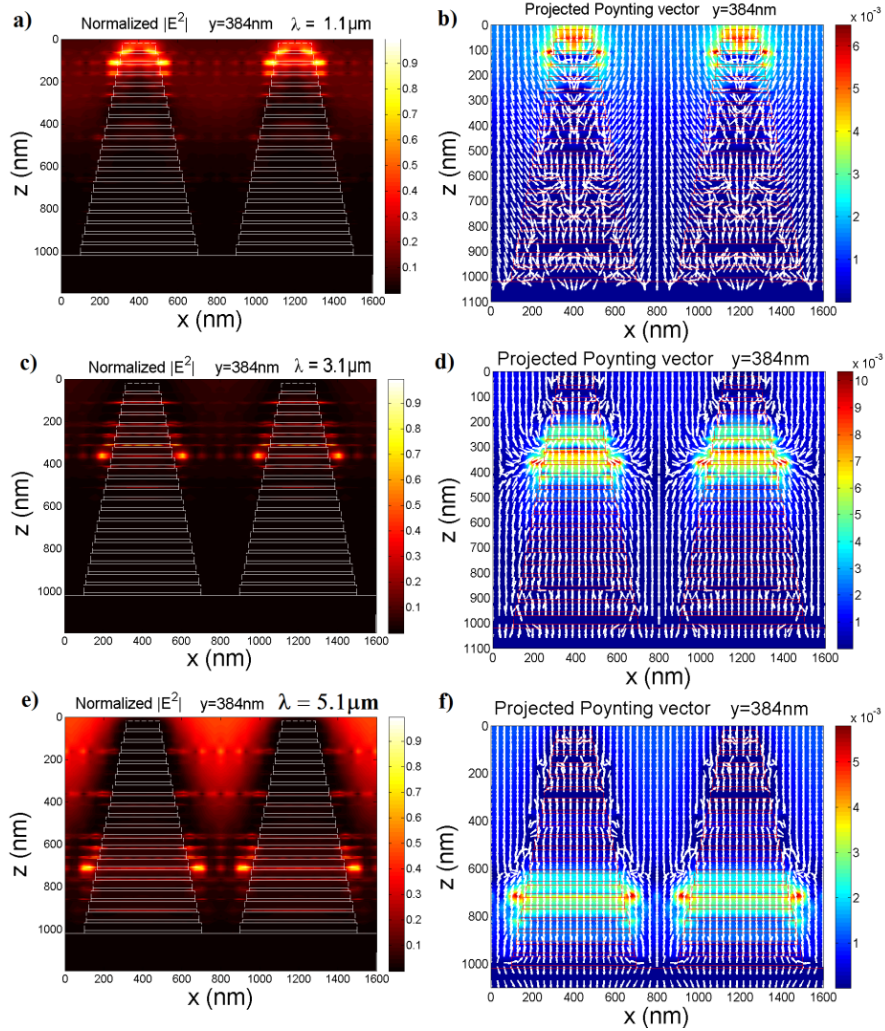

Fig. 8 Map of the field intensity (normalized $|E|^2$) (a, c, e) and of the projected Poynting vector in XZ plane (b, d, f) at incident wavelengths equal to $\lambda = 1.1$ µm, $\lambda = 3.1$ µm and $\lambda = 5.1$ µm for the starting structure. The XZ plane is taken in the middle of the pyramid (y=384 nm).

As the vertical spacing between two stacked resonators increases, i.e. the thickness of the dielectric spacer increases, the plasmonic modes stop hybridizing and recover their original isolated characteristics (Fig. 5). This explains why absorption increases at wavelengths above 6 µm for increasing dielectric spacer thickness but decreases in the 1-2 µm wavelength range (Fig. 4(a)). The dependence of absorption on spacer thickness gives an added tunability to the device, i.e. a way to control mode hybridization. Regarding lateral coupling, a similar effect appears: coupling is reduced between resonators when the lateral period increases (Fig. 3). Moreover, the surface filling fraction $f_s = (L/P)^2$ decreases as the lateral period increases. This means that the part of the unit cell responsible for the absorption (the MMA of size *L*) is proportionally lowered while the proportion of the gold thin film (substrate) becomes more important. As a result, a larger fraction of the incoming light is reflected by this gold substrate, hence lowering the absorption as the period becomes larger (Fig. 3).

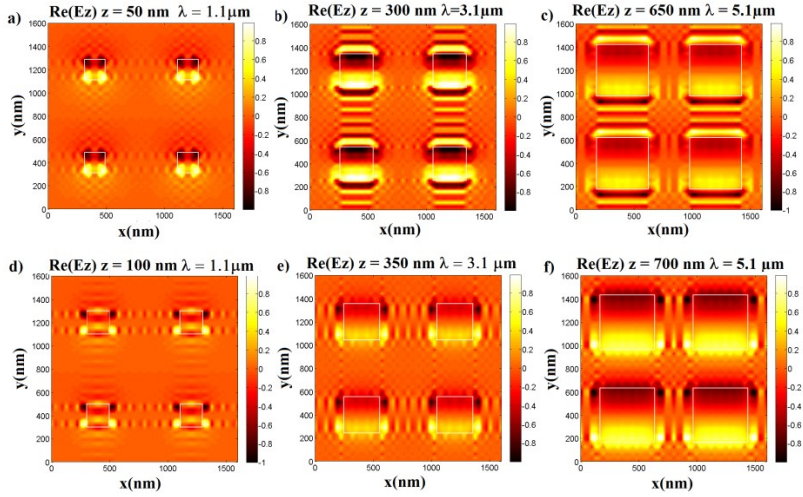

Fig. 9 Real component of $E_z$ in *xy* plane above two successive Au resonant layers (top layers a-c, bottom layers d-f) at different wavelengths: $\lambda$ = 1.1 µm, $\lambda$ = 3.1 µm and $\lambda$ = 5.1 µm.

The intensity map in a vertical cross-section (*xz*) plane (Fig. 8 (a), (c) and (f)) and the Poynting vector flow (Fig. 8 (b), (d) and (f)) confirm that a specific layer of the MMA is resonant and absorbs at a given wavelength. This behavior is observed at three different wavelengths, i.e. below Ge bandgap ($\lambda$=1.1 µm), above Ge bandgap ($\lambda$=3.1µm), and around the upper edge of the operational bandwidth ($\lambda$=5.1µm). Moreover, field intensity is greatly enhanced at the edges of the resonant layer and vertical coupling can be observed through the localization of the intensity on the nearby upper and lower layers. We note that the intensity does not vanish in the dielectric spacers, which enables vertical coupling between metallic layers. These observations are consistent with the associated map of the modulus of the projected Poynting vector in *xz* plane. Superimposed on the map, the arrows represent the direction of the normalized vector. The power flux is clearly disturbed by the resonant layers, creating a vortex-like behavior, which increases the localization of the absorption. At $\lambda$=5.1µm, the higher field intensity in the air is related to the lowered absorption at this wavelength (92%, Fig. 1(b)), and the fact that reflection has to increase concomitantly. In order to determine which mode is excited, we examined electric field maps in horizontal cross-section (*xy*) plane. Maps of the real part of $E_z$ are shown in *xy* planes just above two successive Au layers (i.e. in dielectric spacers), which are resonantly coupled (top layer Figs. 9 (a)-(c), bottom layer Figs.9 (d)-(f)). We can clearly identify dipolar modes that are excited in these two planes. Lateral coupling can be observed since dipoles are aligned and the electric field is non-vanishing between adjacent layers, especially in the bottom of the MMA, where the interdistance between layers is smaller.

Regarding polarization independence (Fig 1(d)), we note that LSP modes of stand-alone metallic square of sub-wavelength size are polarization independent in the *xy* plane (Fig. 5). Since the MMA consists of a stack of metallic squares, there is no strong constraint on the incident polarization state. In addition, the size of the pyramid, as well as the grating period, are small when compared to wavelengths within the OBW. Diffraction is therefore negligible and the whole device acts as an effective (meta-)material containing sub-wavelength metallic particles [7,8].

Finally, we have also verified that the ultra-broadband and highly absorbing properties of the MMA are preserved with crystalline silicon as semi-conductor medium instead of germanium. Silicon is interesting due to lower fabrication costs and larger bandgap. With a lateral period *P = 610* nm, a dielectric thickness $t_{diel}$ = *35* nm and complex refractive index taken from [23], while other parameters are kept the same as in section 2, we numerically found that absorption is also very high on a broad range (Fig 1(c)). Between $\lambda_i$ = 0.2 µm and $\lambda_f$ = 5.7 µm, $\eta_{OBW}$ = 93.8% assuming 70% threshold condition. This results in an operational bandwidth *OBW* = *5.5* μm without any further optimization. The robustness of the ultra-broadband absorption with respect to different dielectric spacer materials is explained by the dependence of the surface plasmons on the surrounding dielectric medium (spacers). If the dielectric permittivity of the background medium is lowered from that of Ge to Si, the absorption peaks are blue-shifted. This effect accounts for slightly reduced OBW obtained with Si instead of Ge. Both Ge and Si exhibit a bandgap, at 1.8μm and 1.1μm respectively. Part of the incoming radiation is then absorbed in the spacer medium as soon as the incident light has wavelength components lower than the bandgap wavelength. However, as seen in Fig. 8(a) for Ge, at λ=1.1μm, the above-explained physical mechanism (plasmon hybridization) is preserved even when the dielectric spacer is absorbing. This shows the importance of using metamaterials, i.e. a combination of dielectric and metallic elements, in order to enable high and broadband absorption over a spectral range, which would be impossible with those elements taken separately. We finally note that the absorbed energy should be dissipated in some way. By virtue of Kirchoff's law, thermal emission will be reinforced if a perfect absorber is designed.

## 5. Conclusions

We have numerically demonstrated the performance of an ultra-broadband pyramidal metamaterial absorber with the exploitation of plasmon hybridization. Hybridization involves localized surface plasmons at the metal-dielectric interfaces. Each layer acts as an individual resonator at frequency related to the aspect ratio of the square layers. We found that dipolar modes are prominent over other high order modes. Electromagnetic coupling through plasmonic hybridization between alternating layers enables the ultra-broadband absorption. The proposed Au/Ge device is able to absorb light over a wavelength range as broad as 5.6 μm (between 0.2 μm and 5.8 μm), with an integrated absorption (over the operational bandwidth) as high as 98.0%. Versatility of the device is highlighted by replacing Ge with Si as spacer material. Moreover, the device shows good angular and polarization insensitivities, at up to 75°. The proposed pyramidal structure allows also the tuning of the operational bandwidth by changing the aspect ratio of the individual resonators. Optimal values of dielectric spacer thickness and lateral period were found and their influence on the absorption properties was discussed.

The reported developments and optimizations are expected to be particularly relevant for experimental realization of the proposed MMA design. In particular, versatility of dielectric material will aid in the development of such a device. The primary challenge would be in the production of the number of layers required to fill the bandwidth over a broad spectrum in a highly controlled manner. However, such a device could be realized with the use of current lithographic techniques such as electron beam lithography, or possibly with nano-imprint lithography, as a way to reduce lengthy processes, while maintaining fidelity in reproduction steps. The substrate, for example, can be made of a variety of material as a way to further reduce cost in production.

### Acknowledgments

The authors would like to thank Stéphane-Olivier Guillaume for his help in DDA simulations. This research used resources of the "Plateforme Technologique de Calcul Intensif (PTCI)" (http://www.ptci.unamur.be) located at the University of Namur, Belgium, which is supported by the F.R.S.-FNRS. The PTCI is member of the "Consortium des Équipements de Calcul Intensif (CÉCI)" (http://www.ceci-hpc.be).